\documentclass[prl,twocolumn,12pts,preprintnumbers,amsmath,amssymb,superscriptaddress,notitlepage]{revtex4}
\usepackage[T1]{fontenc}
\usepackage{textcomp}
\usepackage{amsmath}
\usepackage{amssymb}
\usepackage{graphicx} 
\usepackage[pdftex]{color}
\usepackage{amsfonts}
\usepackage{textcomp}
\usepackage{gensymb}
\usepackage{epsfig}
\usepackage{bm}

\begin{document}

\title{Chiral Majorana Interference as a Source of Quantum Entanglement}

\author{Luca Chirolli$^\star$}
\affiliation{Fundaci\'on IMDEA Nanoscience, E-28049 Cantoblanco Madrid, Spain}
\email{luca.chirolli@imdea.org}

\author{Jos\'e Pablo Baltan\'as}
\author{Diego Frustaglia}
\affiliation{Departamento de F\'isica Aplicada II, Universidad de Sevilla, E-41012 Sevilla, Spain}

\begin{abstract}
Interferometry is a powerful tool for entanglement production and detection in multiterminal mesoscopic systems. 
Here we propose a setup to produce, manipulate and detect entanglement in the electron-hole degree of freedom 
by exploiting  Andreev reflection on chiral one-dimensional channels via interferometry. We study the best possible 
case in which two-particle interferometry produces superpositions of maximally entangled states. This is achieved 
by mixing chiral Dirac channels through chiral Majorana modes. We show that it is possible to extract entanglement 
witnesses through current cross-correlation measurements.
\end{abstract}

\maketitle

{\it Introduction --} 
Entanglement is at the core of Quantum Theory and represents a key resource for quantum information and computation. 
Generation, manipulation, and detection of entangled electrons is at the basis of quantum computing with integrated 
solid-state devices. Among many possibilities, a great attention has been devoted to entanglement in multiterminal 
mesoscopic conductors (see Refs.~\cite{Beenakker-PHentanglement,BurkardRev} for a review) and  the most noticeable 
schemes rely on Cooper pair emission from superconducting contacts \cite{RecherPRB2001,Lesovik}, correlated 
electron-hole entangled states by tunnel barriers \cite{BeenakkerPRL2003}, and integrated single-particle emitters 
\cite{SplettstoesserPRL2009}.

Here, we suggest to produce entangled electron-hole (e-h) pairs through Andreev reflection on chiral electronic states 
\cite{Hoppe,Rickhaus2012,ClarkeNP2014,LeeNP2017} by promoting the e-h degree of freedom (DoF) to an internal 
interferometric state, analogous to spin in co-propagating spin-resolved edge states in the IQHE \cite{Karmakar,
ChirolliDattaDas,ChirolliPRL2013,Karmakar2015}. In a chiral channel, electrons and holes co-propagate. A particle 
state that scatters off a superconductor (SC) results in a coherent e-h superposition that co-propagate on the same 
channel, allowing for e-h interferometry in a controlled way. Interferometry requires a set of linear elements such as 
beam-splitters (BSs) and phase shifters as building blocks for single- and two-particle elements, such as a Mach-Zehnder 
(MZ) and a Hanbury-Brown-Twiss (HBT) interferometer. These elements make possible to generate, manipulate and 
detect entanglement in the {\it e-h} and {\it channel} DoF.

Recent experiments with chiral one-dimensional (1D) channels in contact with $s$-wave superconductors 
opened the way to exploring Andreev reflection on 1D chiral channels \cite{Rickhaus2012,LeeNP2017}. At the same time, 
no proposal for controlled BS, MZ or HBT e-h interferometer is currently available with ordinary superconductors.
The situation is different for topological superconductors (TSCs) hosting chiral Majorana modes at their boundary 
\cite{Beenakker-MajoranaRev}, where interferometric elements were first introduced in Refs.~\cite{FuKanePRL2009,
AkhmerovPRL2009}. Remarkably, in these systems two-particle interferometry produces superpositions 
of maximally entangled states. In particular, as pointed out in Ref.~\cite{StrubiPRL2011}, post-selecting 
states with one fermion per lead yields maximally entangled pairs in the electron-hole space. This would be a quite 
fortunate coincidence in an ordinary interferometer characterized by arbitrary transmission and reflection coefficients. 
The advantage of using chiral Majorana channels is that the production of exactly maximally entangled states
is guaranteed.

Detection of entangled fermions in the context of quantum transport was first proposed as a particular consequence 
of anti-bunching in current cross-correlation measurements for a subclass of states (spin-entangled particles propagating 
along different channels) by using a BS analizer  \cite{BurkardPRB2000,TaddeiPRB2002,BurkardPRL2003}. This was later 
generalized to the case of multiple mode and occupancy entanglement \cite{GiovannettiPRB2006,GiovannettiPRB2007,Baltanas2015}, 
whereby current cross-correlations can provide entanglement witnesses \cite{Horodecki,Lewenstein}. Here, we review these 
proposals in view of single- and two-particle interferometry via chiral Majorana modes and introduce a fundamental phase gate 
that allows to implement a phase shift between electron and hole states at zero energy. The combined action of a ${\mathbb Z}_2$ 
MZ and a phase gate allows for arbitrary rotations in the e-h DoF to perform local operations at will on each propagating 
channel, boosting the entanglement witness power. Our approach makes it possible to exploit single- and two-particle interferometry 
in the e-h DoF as a platform for quantum computation in dual-rail architectures \cite{Chuang,Knill}.

{\it The system --} 
The main ingredients are chiral Dirac modes ($\chi$DMs) and chiral Majorana modes ($\chi$MMs) in quantum 
anomalous Hall insulator/SC structures, as those proposed in Refs.~\cite{FuKanePRL2009,AkhmerovPRL2009}. 
The recent experimental detection of $\chi$MMs in these systems \cite{Zhang} makes the present proposal feasible 
and particularly appealing. The system consists in the two-dimensional (2D) surface of a topological insulator (TI) 
on top of a substrate divided in ferromagnetic (FM) and SC regions. The TI surface hosts a single 2D fermionic 
Dirac cone described by $H_0=-iv(\nabla\times{\bf s})_z$, with ${\bf s}$ a vector of spin Pauli matrices. A FM domain 
wall acts as $H_{\rm fm}=M({\bf r})s_z$ and gaps the system everywhere apart from a line where the domain wall 
changes sign. Along this line a 1D $\chi$DM forms, analogous to the edge states of the IQHE, and 
can be used as an electronic wave guide. Similarly, SC proximity induces singlet pairing described by 
$H_{\rm sc}=\Delta \sum_{{\bf k}}c^\dag_{{\bf k},\uparrow}c^\dag_{-{\bf k},\downarrow}+{\rm H.c}$ 
that opens a topological gap, thus realizing a 2D TSC. Gapless $\chi$MM form along the border between 
the SC and the FM regions. By properly arranging these regions it is possible to realize an interferometric setup composed 
by several linear elements. We now characterize the transport in terms of scattering matrices in the Landauer-B\"uttiker 
formalism adapted to describe $\chi$MMs \cite{LiButtikerPRB2012}. 

\begin{figure}
  \centering
  \includegraphics[width=0.5\textwidth]{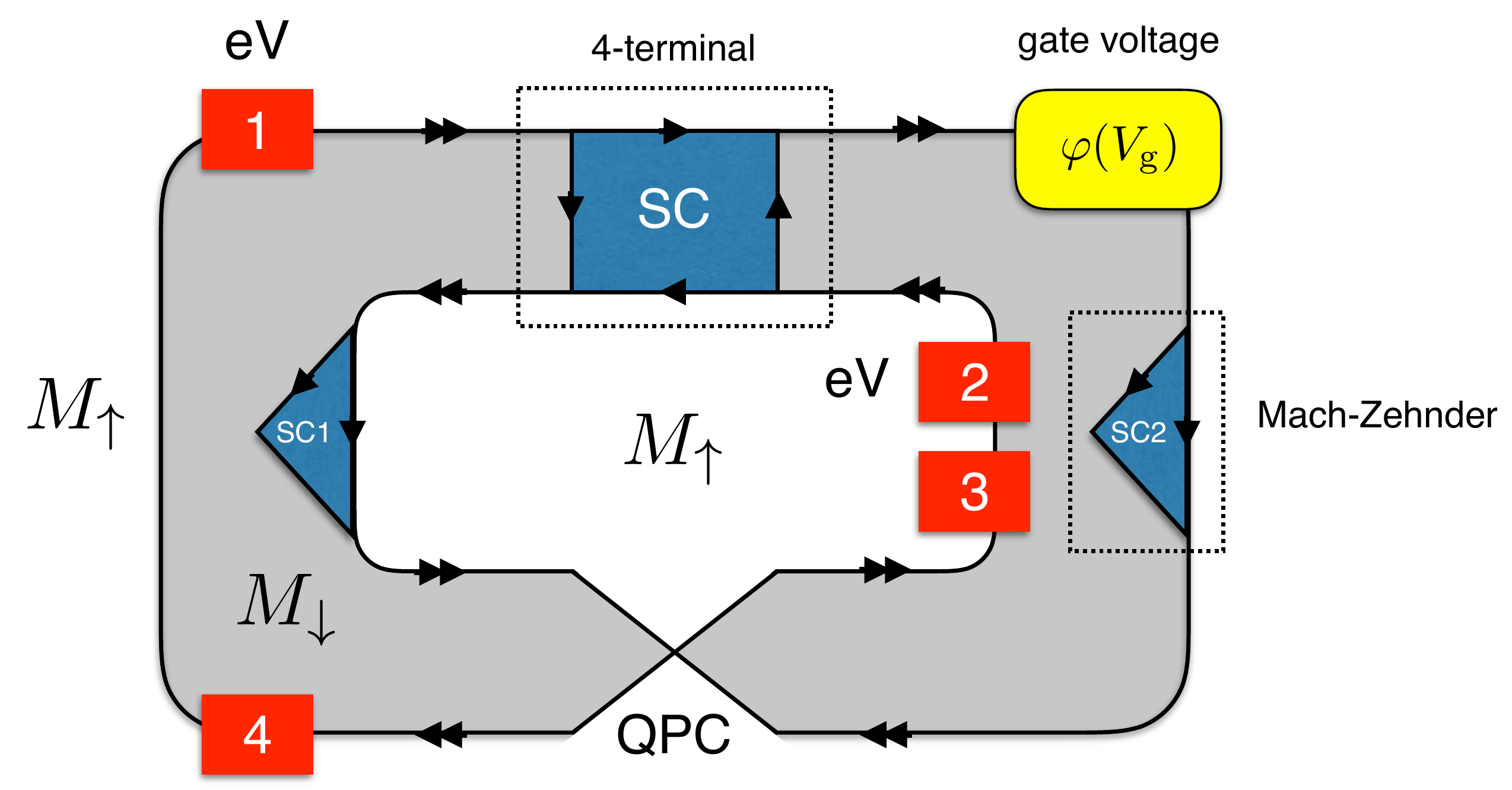}
  \caption{Setup: a domain wall on top of a TI 2D surface state generates $\chi$DMs (double arrows) and an 
  $s$-wave SC opens a topological gap, at whose boundary with the domain wall $\chi$MMs form (single arrow). 
  Carriers are injected into the system by biasing contacts 1 and 2 and go first through the four-terminal device, 
  that acts as a two-particle interferometer.  The outgoing carriers then undergo local operations through phase 
  shifts and MZs and collide into a QPC that mixes the channels. Currents and correlations are measured in 
  contacts 3 and 4. The MZs and phase shifts can be also placed after the QPC. }
  \label{Fig:setup}
\end{figure}

{\it The interferometric setup --}
The setup is illustrated in Fig.~\ref{Fig:setup}. We follow the notation of Ref.~\cite{StrubiPRL2011} and denote $\chi$DMs 
with double arrow lines and $\chi$MMs with single arrow lines.  Given that $\chi$DMs have $\langle s_y\rangle=-1$, we can 
regard the fundamental excitations as spinless Dirac fermions described by fermionic operators $a(\epsilon)$ at energy 
$\epsilon$. For energies $0<\epsilon\ll \Delta$ we define electron- and hole-like states in channel $i$ at energy $\epsilon$ 
as $a_{i,+}(\epsilon)=a_i(\epsilon)$ and $a_{i,-}(\epsilon)=a_i^\dag(-\epsilon)$, and introduce a e-h 
DoF $\tau=\pm 1= e,h$. Analogously, at the boundary of the TSC a single $\chi$MM flows, either clockwise  
$\gamma_1(\epsilon)$ at the boundary between the TSC and the $M_\uparrow$ magnetic domain, or 
anti-clockwise $\gamma_2(\epsilon)$  at the boundary between the TSC and the $M_\downarrow$ magnetic domain. 

Current is injected in the system upon biasing contacts 1 and 2 and it is collected in contacts 3 and 4, that are kept grounded. 
The resulting current and noise in contact 3 and 4 can be obtained in terms of unitary scattering matrices $S$ relating incoming 
Dirac modes $a_{j,\sigma}$ to outgoing Dirac modes $b_{i,\tau}$
\begin{equation}
b_{i,\tau}(\epsilon)=\sum_{j=1,2; \sigma=\pm}S_{i,\tau;j,\sigma}(\epsilon)a_{j,\sigma}(\epsilon).
\end{equation}
Particle-hole symmetry implies $S(\epsilon) =\tau_x S^*(-\epsilon)\tau_x$. 

The setup in Fig.~\ref{Fig:setup} 
is characterized by four kinds of linear elements: i) a BS, ii) a MZ interferometer, iii) a phase shifter, and iv) a four terminal element.  
Additionally, an ordinary quantum point contact (QPC) is introduced to mix $\chi$DMs, in analogy with interferometry in the IQHE edge 
states \cite{Neder2007}. Some of these elements have been suggested in the literature, so that we here review them in the spirit 
of two-particle interferometry.

{\it Beam splitter --}
In this context, we denote a BS as an element that takes an incoming $\chi$DM and produces two outgoing $\chi$MMs. The tri-junction 
between a magnetic domain wall and a TSC forces the incoming particle $a_+$ and hole $a_-$ states to split into two $\chi$MMs, 
$\gamma_1$ and $\gamma_2$ \cite{FuKanePRL2009,AkhmerovPRL2009}. For energies much smaller than the SC gap, we can 
assume the BS scattering matrix to be energy independent, reading \cite{FuKanePRL2009,AkhmerovPRL2009}
\begin{equation}
\left(\begin{array}{c}
\gamma_1(\epsilon)\\
\gamma_2(\epsilon)
\end{array}\right)=\frac{1}{\sqrt{2}}\left(
\begin{array}{cc}
1 & 1\\
i & -i
\end{array}\right)\left(\begin{array}{c}
a_+(\epsilon)\\
a_-(\epsilon)
\end{array}\right).
\end{equation}

{\it Mach-Zehnder -- }
By combing two BSs separated by a region where the two $\chi$MMs propagate along two paths of different length one can realize 
a MZ interferometer for electrons and holes described by \cite{FuKanePRL2009,AkhmerovPRL2009,Alos-Palop}
\begin{equation}
\left(\begin{array}{c}
a_+\\
a_-
\end{array}\right)_R=S_{\rm bs}^\dag\left(
\begin{array}{cc}
e^{i\pi n_v+ikL_1} & 0\\
0 & e^{ikL_2}
\end{array}\right)S_{\rm bs}\left(\begin{array}{c}
a_+\\
a_-
\end{array}\right)_L,
\end{equation}
where $k(L_1-L_2)=\epsilon\delta L/v_M$ is the phase difference gathered at energy $\epsilon$ and $n_v$ is the number of 
vortices in the SC, with $v_M$ the velocity of the $\chi$MMs. The scattering matrix thus mixes incoming chiral electron and hole states 
in the left ($L$) lead to outgoing chiral electron and hole states in the right ($R$) lead. At finite energy, by varying the path length difference 
$\delta L$ any arbitrary scattering matrices between incoming and outgoing states in a given channel can be generated. At zero 
energy this element is expressed as a $\tau_x$ scattering matrix in the e-h space, that represents a ${\mathbb Z}_2$ MZ 
interferometer being able only to change a particle into a hole, and viceversa. 

{\it Phase shifter --}
A fundamental ingredient appearing in the setup of Fig.~\ref{Fig:setup} is the phase shifter between electrons and holes in a given 
Dirac channel. This can be easily accomplished by a top gate that locally shifts the chemical potential. For a gate voltage $V_{\rm g}$ 
such that $|V_{\rm g}|\ll M$, where $M$ is the magnitude of the Zeeman splitting of the magnetic domains, electrons and holes will in 
general acquire an opposite phase,
\begin{equation}
a_{+}(\epsilon)\to e^{i\varphi_g}a_+(\epsilon),\qquad a_{-}(\epsilon)\to e^{-i\varphi_g}a_-(\epsilon),
\end{equation}
with $\varphi_g=(e/v)\int^L_0dxV_g(x)$. Importantly, 
this phase is energy-independent so that also carriers at $\epsilon=0$ acquire it. The scattering matrix associated to the phase shift 
is $P=\exp(i\varphi_g\tau_z)$. This element is of paramount importance in that, combined with the ${\mathbb Z}_2$ MZ, it provides 
a way to rotate the states in the e-h space and generate any superposition state. Moreover, a channel-dependent shift can be obtained 
by modifying the path length of a given channel. This can be achieved by moving the domain wall through a magnetic field.

{\it Four-terminal element --}
Finally, the core of the setup in Fig.~\ref{Fig:setup} is a four terminal device that mixes two incoming $\chi$DMs into two outgoing $\chi$DMs. 
In terms of electron and hole channels, the element mixes four incoming states into four outgoing states. This four-terminal element was 
introduced in Refs.~\cite{FuKanePRL2009,StrubiPRL2011} and it is described by the scattering matrix 
\begin{equation}\label{Shbt}
\left(\begin{array}{c}
b_{1+}\\
b_{1-}\\
b_{2+}\\
b_{2-}
\end{array}\right)=\frac{1}{2}
\left(\begin{array}{cccc}
1 & 1 & 1 & 1\\
1 & 1 & -1 & -1\\
1 & -1 & -\eta & \eta\\
1 & -1 & \eta & -\eta
\end{array}\right)
\left(\begin{array}{c}
a_{1+}\\
a_{1-}\\
a_{2+}\\
a_{2-}
\end{array}\right),
\end{equation}
with $\eta=(-1)^{n_v}e^{i\epsilon\delta L/\hbar v_M}$ a phase due to the propagation of the $\chi$MMs in the interferometer 
(here $\delta L$ is the path length difference between the two-particle trajectories \cite{FuKanePRL2009,StrubiPRL2011}). 
At zero energy the phase can be only $\eta=\pm 1$, depending on the number of vortices $n_v$ in the system. A more 
generic four-terminal element can be obtained by allowing for a cross talk between the $\chi$MMs in the SC region. 

{\it Cross-correlations as Entanglement witness --} 
The general idea developed in Refs.~\cite{BurkardPRB2000,TaddeiPRB2002,BurkardPRL2003,GiovannettiPRB2006,
GiovannettiPRB2007} is that, given an unknown initial state that is possibly entangled in the e-h and channel DoF, 
it is possible to establish the presence of entanglement via measuring current cross-correlations after mixing the 
channels through a QPC and relate the cross-correlator to an entanglement witness. We now show that witnessing 
entanglement in a e-h system is not only possible, but also much more effective, thanks to the possibility to insert MZs 
before and after the QPC. 

The most general unknown two-particle state at the input of the QPC can be cast in the generic form
\begin{equation}\label{PsiHBTgeneric}
|\Psi\rangle=\sin\theta(\cos\phi|\Phi_{11}\rangle+\sin\phi|\Phi_{22}\rangle)+\cos\theta|\Phi_{12}\rangle,
\end{equation}
with $\theta,\phi\in[0,\pi/2]$ and $|\Phi_{ij}\rangle$ two particle states at energy $\epsilon$ in lead $i$ and $j$,   
$|\Phi_{ij}\rangle=\sum_{\alpha,\beta}\Phi^{(ij)}_{\alpha,\beta}a^\dag_{i,\alpha}(E)a^\dag_{j,\beta}(E)|0\rangle$, 
where $|0\rangle$ is the grounded Fermi sea. The states satisfy $\Phi^{(jj)}_{\alpha,\beta}=-\Phi^{(jj)}_{\beta,\alpha}$ 
and the normalization conditions $\sum_{\alpha,\beta}|\Phi^{12}_{\alpha,\beta}|^2=1$ 
and $\sum_{\alpha,\beta}|\Phi^{(jj)}_{\alpha,\beta}|^2=1/2$. The state described by Eq.~(\ref{PsiHBTgeneric}) displays 
entanglement in the {\it occupation number} and 
{\it channel} DoF \cite{SchuchPRL2004,WisemanPRL2003}. 

Before the QPC, we induce a phase difference between the electron and hole in each channel by local gate voltages and 
domain wall displacement. For the moment we do not consider the MZs that are present in the setup of Fig.~\ref{Fig:setup}. 
The QPC mixes the channel of the incoming particles without changing the electron/hole character of the particle injected. 
The outgoing states after the combined system phase shifter plus QPC is given by $b_{j,\tau}=\sum_{j'=1,2}S^\tau_{j,j'}a_{j',\tau}$, 
where $b_{j,\tau}$ are the outgoing states after the QPC in lead $j$ with electron/hole character $\tau$, where the scattering 
matrix is given by
\begin{equation}
S^{\tau}=\left(\begin{array}{cc}
r & t'e^{i\varphi_\tau}\\
t & r'e^{i\varphi_\tau} 
\end{array}\right).
\end{equation}
The phase $\varphi_\tau$ accumulated before the QPC has two contributions, the gate contribution $\varphi_g$, that is opposite 
for electrons and holes, and a dynamical phase difference due to the different path length between the four-terminal scattering 
region and the QPC along the two possible paths. This contribution is the same for particles and holes, so that we can write 
$\varphi_\tau=\varphi_L+\tau\varphi_g$, with $\varphi_L=\epsilon \delta L/v$ and $v$ the velocity of Dirac modes. 
We now consider the dimensionless current cross correlator between the output channels 3 and 4,
\begin{equation}
C_{34}\equiv \frac{h^2\nu^2}{2e^2}\lim_{{\cal T}\to \infty}\int_0^{\cal T}\frac{dt_1dt_2}{{\cal T}^2} \langle I_{3}(t_1)I_{4}(t_2)\rangle,
\end{equation}
where ${\cal T}$ is the measurement time. 
The current operator of chiral fermions in lead $i$ is written in terms of particle and hole contributions as
\begin{equation}\label{Eq:currentOp}
I_i(t)=\frac{e}{h \nu}\sum_{\epsilon,\omega,\tau=\pm}e^{-i\omega t}\tau~ b^\dag_{i,\tau}(\epsilon)b_{i,\tau}(\epsilon+\hbar \omega),
\end{equation}
and the average $\langle\ldots\rangle$ is taken over the incoming state by assuming a discrete spectrum characterised by a density 
of states $\nu$ in each lead.  Importantly, electrons and holes contribute with different sign to the current, that is accounted for by the 
$\tau$ in Eq.~(\ref{Eq:currentOp}). In each lead there are incoming and outgoing states. However, due to the chirality of states localized 
at the domain wall boundary, there is no back scattering and the current can be described only in terms of outgoing channels. 

The quantity $C_{34}$ has the advantage of being a linear function of the input state \cite{GiovannettiPRB2007}. Assuming a  
QPC characterized by $r'=r=\sqrt{1-T}$ and $t'=t=i\sqrt{T}$, with $T$ the transmission probability of the QPC  we find
\begin{eqnarray}
C_{34}(\Psi)&=&T(1-T)\left[-\sin^2\theta+v\sin^2\theta\sin(2\phi)+w\cos^2\theta\right]\nonumber\\
&+&\frac{1}{2}\cos^2\theta\sum_{\tau\sigma}\tau\sigma|\Phi^{12}_{\tau,\sigma}|^2,
\end{eqnarray}
where $v$ and $w$ are real quantities satisfying $|v|,|w|\leq 1$ that can be expressed in terms of the phases $\varphi_{\tau}$ as
\begin{eqnarray}
v&=&2{\rm Re}\sum_{\tau,\sigma}(\Phi^{(11)}_{\tau,\sigma})^*\Phi^{(22)}_{\sigma,\tau}e^{i\varphi_\tau+i\varphi_\sigma},\\
w&=&\sum_{\tau,\sigma}\sigma\tau(\Phi^{(12)}_{\tau,\sigma})^*\Phi_{\sigma,\tau}^{(12)}e^{i(\varphi_\tau-\varphi_\sigma)}.
\end{eqnarray}
The correlator $C_{34}$ is very similar to that of Ref.~\cite{GiovannettiPRB2007}. Antisymmetry of the $|\Phi_{jj}\rangle$ states implies 
that $\varphi_g$ drops from $v$.  Analogously, the phase $\varphi_L$ drops from $w$. By further redefining $\varphi_g\to \pi/2+\varphi_g$ 
we have that $w=\sum_{\tau,\sigma}(\Phi^{(12)}_{\tau,\sigma})^*\Phi_{\sigma,\tau}^{(12)}e^{i\varphi_g(\tau-\sigma)}$. 

As a particular case we consider incoming states with singly occupied channels 1 and 2 by choosing $\theta=0$ in the generic input 
state (\ref{PsiHBTgeneric}). The current correlator is found to be
\begin{equation}\label{Eq:C34-12}
C_{34}(\Phi_{12})=\frac{1}{2}\sum_{\tau\sigma}\tau\sigma|\Phi^{12}_{\tau,\sigma}|^2+T(1-T)w,
\end{equation}
The quantity $w$ captures all the relevant information on the input state. One can show that $w$ is non-negative 
for separable input states \cite{GiovannettiPRB2006}. The first term in Eq.~(\ref{Eq:C34-12}) is nothing but the cross correlator before the 
QPC, $C_{12}(\Phi_{12})=\frac{1}{2}\sum_{\tau\sigma}\tau\sigma|\Phi_{\tau\sigma}^{(12)}|^2$. Experimentally one can act on the QPC and 
switch the tunneling between counter propagating states on and off (by setting $T=0$ or $T=1$) and measure separately $C_{34}(\Phi_{12})$ 
and $C_{12}(\Phi_{12})$. It then follows that the case $C_{34}(\Phi_{12})-C_{12}(\Phi_{12})<0$ witnesses the presence of entanglement in the 
state $|\Phi_{12}\rangle$. 

For $\theta\neq 0$, i.e., incoming channels 1 and 2 with fluctuating local occupancy in (\ref{PsiHBTgeneric}), $C_{34}(\Psi)$ can be 
related to the entanglement of formation $E_f(\Psi)$ \cite{Bennett}: generalized Werner states \cite{Vollbrecht,Werner} are defined by 
introducing a joint orthonormal basis for ports 1 and 2 formed by states $|\chi_k\rangle_1\otimes|\chi_{k}\rangle_2$ and 
$|\Psi^{(\pm)}_{kk'}\rangle=(|\chi_k\rangle_1\otimes|\chi_{k'}\rangle_2\pm|\chi_{k'}\rangle_1\otimes|\chi_{k}\rangle_2)/\sqrt{2}$ 
with $k<k'$, where $k$ enumerates all configurations with two or fewer particles per port. It then follows that the entanglement of
formation of a state $\rho$ can be lower bounded by the quantity $W(\rho)=\sum_{kk'}\langle\Psi^{(-)}_{kk'}|\rho|\Psi^{(-)}_{kk'}\rangle/2$. 
Analogously, we can relate the net correlator $\delta C_{34}(\Psi)\equiv C_{34}(T=1/2)-C_{34}(T=0)=C_{34}(\Psi)-C_{12}(\Psi)$, 
which depends only on the quantities $v$, $w$, and the angle $\theta$, to a lower bound to the entanglement of formation through
\begin{equation}
W(\Psi)=-2\delta C_{34}(\Psi)+\cos^2(\theta)/2.
\end{equation}
By noticing that $W(\Psi)>-2\delta C_{34}(\Psi)$ we find that $W(\Psi)>1/2$ and, consequently, $E_f(\Psi)>0$ whenever 
$\delta C_{34}(\Psi)<-1/4$ (see Refs. \cite{GiovannettiPRB2006,GiovannettiPRB2007}). Thus, the sign of $-2\delta C_{34}(\Psi)-1/2$ 
is sufficient to witness the presence of entanglement in the initial state $\Psi$. We can further post process the data and obtain full 
information about the state. First of all, we use that $v(\varphi_L+\pi/2)=-v(\varphi_L)$. This allows to define 
$\delta C_{34}^{(\pm)}=(\delta C_{34}(\varphi_L+\pi/2) \pm \delta C_{34}(\varphi_L))/2$ such that 
\begin{eqnarray}
\delta C_{34}^{(+)}(\varphi_g)&=&(w(\varphi_g)\cos^2\theta-\sin^2\theta)/4,\\
\delta C_{34}^{(-)}(\varphi_L)&=&v(\varphi_L)\sin^2\theta\sin(2\phi)/4.
\end{eqnarray}
We then notice that $\cos^2\theta=2C_{12}(\Psi)/(2\bar{w}-1)$, where $\bar{w}=\int \frac{d\varphi_g}{2\pi}w(\varphi_g)$. Upon introducing 
$\delta\bar{C}^{(+)}_{34}=\int \frac{d\varphi_g}{2\pi}\delta C^{(+)}_{34}(\varphi_g)=((1+\bar{w})\cos^2\theta-1)/4$, we can express
\begin{equation}
\cos^2\theta=\frac{2}{3}\left(4\delta\bar{C}_{34}^{(+)}-C_{12}+1\right),
\end{equation} 
allowing to establish the occupation and channel admixture as a function of measurable quantities. Finally, we notice that the states 
$|\Phi_{jj}\rangle$ can only be e-h singlets, so that $v$ depends only on the relative phase difference between $\Phi_{11}$ and $\Phi_{22}$. 
By varying $\varphi_L$ one can then access $\sin(2\phi)=2({\rm max}_{\varphi_L}\delta C_{34}^{-}-{\rm min}_{\varphi_L}\delta C_{34}^{-})/(1-\cos^2\theta)$. 
The analysis allows to fully access the occupation-number (e-h) and channel DoF entanglement by further exploiting the phase before the QPC 
\cite{GiovannettiPRB2007}. This result can be generalized to generic mixed input states $\rho$, as the combination of the 
maps $C_{12}$ and $C_{34}$ preserves the linearity of $\delta C_{34}$.

{\it HBT state -- }
Having established the general entanglement witnessing protocol via current cross-correlation measurements, we now apply it to the output state of the 
two-particle interferometer in the setup of Fig.~\ref{Fig:setup}. Upon biasing only the contacts 1 and 2 the incoming state at energy $\epsilon$ reads 
$|\Psi_\epsilon\rangle_{\rm in}=a^\dag_{1+}(\epsilon)a^\dag_{2+}(\epsilon)|0\rangle$ and the outgoing state reads 
$|\Psi_\epsilon\rangle_{\rm out}=S_{j,\mu;1,+}S_{k,\nu;2,+}b^\dag_{j,\mu}(\epsilon)b^\dag_{k,\nu}(\epsilon)|0\rangle$ (summed over repeated indexes), 
with $S$ the scattering matrix in Eq.~(\ref{Shbt}). In Ref.~\cite{StrubiPRL2011} the system was studied as a Hanbury-Brown-Twiss (HBT) two-particle 
interferometer and it was recognized that post-selecting states with a single fermion per lead yields maximally entangled states. In this case 
the current cross correlations allows to access $v=-{\rm Re}[\eta e^{2i\varphi_L}]$ and $w=\{[1+\cos(2\varphi_g)]+{\rm Re}[\eta][1-\cos(2\varphi_g)]\}/2$, 
and the associated witness for $T=1/2$ and $\eta$ real is 
\begin{equation}
W(\Psi_{\rm HBT})=\frac{1}{8}[3-\eta+2\eta\cos(2\varphi_L)-(1-\eta)\cos(2\varphi_g)]. 
\end{equation}
In particular, for $\eta=-1$, $\varphi_L=\pi/2$ and $\varphi_g=\pi/2$ one can reach $W=1$, that corresponds to $E_f=1$. This means that the phases 
are means to rotate the initial state to have maximum overlap with the generalized Werner states, confirming that the state coming out from the HBT 
interferometer is a superposition of maximally entangled states. 

{\it Local operations --} The quantity $C_{34}$ differs from that of
Ref.~\cite{GiovannettiPRB2007} in the measured observable:
particle current in the original case in contrast to charge current in 
the present case. This grants a much more powerful characterization 
of the incoming states. By inserting MZs before the QPC we can 
rotate the e-h state on each channel and measure any linear 
combination of the three Pauli matrices $\tau_i$, with $i=1,2,3$. 
This operation only affects the state $|\Phi_{12}\rangle$ 
\footnote{The states $|\Phi_{jj}\rangle$ are e-h singlets, so that 
any single-particle rotation can only affect the global phase of the 
state $\Phi_{jj}$.} and $C_{12}$ measurements can assess every 
local single-particle observable. The insertion of MZs and phase 
shifters together with the possibility of switching the QPC on and 
off give us the opportunity to cross-correlate the local operation 
and to perform a full tomography of the input state.

{\it Acknowledgments -- }
The authors are thankful to F. Taddei, V. Giovannetti, and C. W. J. Beenakker for very useful discussions. 
LC acknowledges funding from the European Union's Seventh Framework Programme (FP7/2007-2013) 
through the ERC Advanced Grant NOVGRAPHENE (GA No. 290846) and the Comunidad de Madrid through 
the grant MAD2D-CM, S2013/MIT-3007. DF and JPB acknowledge support from MINECO (Spain) with 
FEDER funds through project No. FIS2014- 53385-P. 

\bibliography{biblio}{}

\end{document}